# Effects of sapphire nitridation and growth temperature on the epitaxial growth of hexagonal boron nitride on sapphire

**Kawser Ahmed [*,1], Rajendra Dahal[1], Adam Weltz[2], James J.-Q. Lu[1], Yaron Danon[2], and Ishwara B. Bhat[1]**

[1] Department of Electrical, Computer, and Systems Engineering, Rensselaer Polytechnic Institute, Troy, NY 12180, USA
[2] Department of Mechanical, Aerospace and Nuclear Engineering, Rensselaer Polytechnic Institute, Troy, NY 12180, USA

This paper reports on the epitaxial growth of hexagonal boron nitride (hBN) films on sapphire substrates in a cold wall chemical vapor deposition (CVD) system where different sapphire nitridation and hBN growth temperatures were employed. A thin and amorphous nitridated layer was formed at a low temperature (850 °C), which enabled subsequent epitaxial hBN growth at 1350 °C. The influences of the sapphire nitridation temperature and the growth temperature on the film quality were analyzed by X-ray diffraction (XRD) measurements. Higher than optimum nitridation and growth temperatures improve the crystalline quality of the nitridated layer, but does not favor the epitaxial growth of hBN. hBN films grown at the optimum conditions exhibit the c-lattice constant of 6.66 Å from the XRD θ-2θ scan, and the characteristic in plane stretching vibration at 1370.5 $cm^{-1}$ from Raman spectroscopy. X-ray photoelectron spectroscopy analysis confirmed the formation of stoichiometric hBN films with excellent uniformity.

**1 Introduction** Hexagonal boron nitride (hBN) is a wide bandgap semiconductor with $sp^2$-hybridized atomic sheets of boron and nitrogen. This material has attracted much attention for its properties such as high resistivity [1, 2], high thermal conductivity (2000 $Wm^{-1}K^{-1}$) [3], and stability in aggressive chemical environments and at high temperatures (up to 1000 °C) [4]. hBN, an insulating isomorph of graphene, has a small (1.7%) lattice mismatch to graphene and is expected to be atomically smooth and free from dangling bonds because of its $sp^2$-hybridized bonding and weak interplanar Van der Waals bond [5]. Hence, hBN is an excellent candidate to be used as a supporting substrate and gate dielectric for graphene based electronics [5]. hBN is also an emerging material for deep UV photonics [1, 2]. hBN has been widely studied for solid state thermal neutron detector application, since $^{10}B$, a constituent element of hBN, has a large thermal neutron capture cross section (3840 barns) [2, 6].

$sp^2$-BN (i.e., hBN and rhombohedral BN, rBN) growth has been performed by chemical vapor deposition (CVD) on different substrates such as Cu [7], Ni [8], c-sapphire [1, 2, 9-12], and 6H-SiC [12]. A controlled growth of hBN on large area substrates with high growth rate is required to fulfill the entire spectrum of potential application of this material. This paper reports on the epitaxial growth of hBN on 2 inch sapphire substrates using a low pressure CVD process. The precursors were triethylboron (TEB) for boron and ammonia ($NH_3$) for nitrogen. Direct CVD growth of single crystal $sp^2$-BN on c-axis oriented sapphire is difficult, if not impossible; previous works showed that only turbostratic BN (tBN), a disordered $sp^2$-BN phase and an intermediate between $sp^2$-BN and amorphous BN (aBN), is formed when directly grown on sapphire [10, 11]. In-plane lattice constants of hBN and sapphire are 2.5 Å and 4.758 Å, respectively, indicating a large lattice mismatch [10]. Some groups [9, 12] reduced this lattice mismatch by introducing an AlN buffer layer (in-plane lattice constant of 3.11 Å [10]), grown by a sapphire nitridation step at the growth temperature ($T_G$). In this work, a thin and amorphous buffer layer, formed by a low temperature (850 °C) sapphire nitridation step, is shown to favor epitaxial hBN growth. The critical process parameters are nitridation temperature ($T_N$) and $T_G$. The influences of these parameters on the quality of the hBN films are investigated here. Sapphire nitridation mechanism was explained by some earlier studies [13, 14]. Adsorbed nitrogen atoms (from $NH_3$ decomposition) replaces oxygen atoms on sapphire to form N-Al bonds [13, 14]. The sapphire surface is converted into amorphous $AlN_xO_{1-x}$ for low $T_N$ and short nitridation duration ($t_N$). Higher $T_N$ and longer $t_N$ result in decreasing N-O bonds and improving crystallinity of the nitridated layer [13]. Crystalline AlN formation can occur for sufficiently high $T_N$ and long $t_N$ [13]. Due to lattice mismatch between the nitridated layer and the sapphire substrate, accumulation of strain energy occurs with long $t_N$, which can cause stress enhanced migration and form protrusions on the nitridated surface to reduce excess energy [14, 16]. Protrusion formation also increases with increasing $T_N$, since stress enhanced migration is a thermally activated process [16]. Microscopically flat nitridated layer on sapphire was found suitable for epitaxial growth of GaN [14, 16] and AlN [15]. For these GaN and AlN growths, flat nitridated layer improved adatom mobility and thus favored 2D growth, while nitridated layer with protrusions suppressed the lateral motion of adatoms and favored 3D growth [14-16]. Similarly, increased lateral growth of hBN

* Corresponding author: e-mail ahmedk2@rpi.edu, Phone: +15189619314

(with larger domains) on nitridated sapphire should be favored if the nitridated layer is atomically flat (i.e., smaller protrusion density) with low $T_N$ and short $t_N$.

**2 Experimental methods** Epitaxial growth of hBN was performed in a cold wall horizontal CVD system with a SiC coated graphite susceptor. A cold wall CVD system offers the advantage of less contamination from the interaction of the precursor vapor and the reactor wall. The reactor wall is water cooled to prevent the deposition on it. The reactor pressure was 100 torr, and hydrogen was used as the carrier gas with a fixed flow rate of 2 slm for the entire growth duration in all the growth experiments. TEB is an organometallic reagent which is liquid at room temperature. Hence, TEB vapor was carried into the reactor by bubbling hydrogen gas through the TEB liquid kept in a stainless steel bubbler at room temperature. Sapphire substrates were nitridated for 10 minutes with a $NH_3$ flux of 55 μmol/s at different temperatures (850 °C, 1100 °C, and 1350 °C) to form a thin buffer layer before the epitaxial hBN growth. hBN growth temperatures varied from 1200 °C to 1500 °C. For all the growth experiments, the growth duration was 40 minutes, and V/III ratio was 300 with the precursor fluxes of 150 μmol/s for $NH_3$ and 0.5 μmol/s for TEB.

X-ray diffraction (XRD) was performed to study the crystalline quality and the c-lattice constant of the hBN films. For XRD measurements, a Bruker D8-Discover X-ray Diffractometer was employed using Cu Kα radiation (with a wavelength of 1.541 Å). Ni filter was used to suppress Cu Kβ photons (i.e., diffraction of Cu Kβ line) from reaching the detector. Raman spectroscopy was used to further investigate the hBN phase of the deposited films. A Witec Alpha 300R confocal Raman imaging system with a 532 nm laser excitation source was used to collect the Raman spectra at room temperature. Elemental composition and chemical bonding states of the hBN films were analyzed by X-ray photoelectron spectroscopy (XPS). XPS measurements were performed using a Phi Versaprobe XPS system with a focused monochromatic Al Kα x-ray source (with an energy of 1486.6 eV). In order to determine the optical quality of the grown hBN films, the absorption spectra of the films were measured using a Varian Cary 6000i UV-vis-NIR spectrophotometer with a wavelength range of 175 nm to 1800 nm. A Dektak 8 surface profiler was used to determine the thickness of the hBN films by measuring the step height between the film surface and the bare sapphire substrate.

**3 Results and discussion** The influence of sapphire nitridation on the crystalline quality of hBN films is shown in Fig. 1. In an XRD θ-2θ scan of a highly crystalline hBN film grown on nitridated sapphire, peaks centered at 26.7° and 41.7° should be present, corresponding to the (002) hBN plane and the (006) sapphire plane, respectively [9]. tBN shows a broad peak at lower angle than 26.7° [9]. Depending on the crystalline quality of the AlN formed by the nitridation step, a peak may or may not be seen at 36°, corresponding to the (002) AlN plane. Figure 1 (a) shows the XRD pattern in the θ-2θ configuration of an hBN film grown directly on sapphire, while Fig. 1 (b), (c), and (d) illustrate the θ-2θ scans of hBN films with $T_N$ values of 850 °C, 1100 °C, and 1350 °C, respectively. $T_G$ was kept fixed at 1350 °C for all the four growths. A broad and weak hBN (002) plane peak was found at around 25.95° for hBN grown without a nitridated layer buffer and it corroborates previous reports [10, 11] that $sp^2$-BN grown on non nitridated sapphire substrates is of poor crystalline quality. Sapphire nitridation at a temperature of 850 °C resulted in a much better film as far as the X-ray diffraction is concerned, with the hBN (002) peak at 26.7° with a full width at half maximum (FWHM) value of 0.25°. This XRD peak refers to the c-lattice constant of 6.66 Å, which is equal to the c-lattice constant measured for bulk hBN [17, 18]. hBN (002) plane peak shifted from 26.7° to 26.3° (with a higher FWHM of 1.7°) when $T_N$ was 1100 °C, and to 25.85° (with a FWHM of 1.7°) when the $T_N$ was 1350 °C. These films can be identified as tBN. Sapphire nitridation process with $NH_3$ in a CVD process was observed for temperatures greater than 800°C [19]. Nitridation at 850 °C should form an amorphous and thin $AlN_xO_{1-x}$ layer which is further confirmed by the absence of any peak around 36° in Fig. 1 (b). Nitridated layer has better crystallinity as the $T_N$ increased, as evidenced from the very broad peak centered around 36° for a $T_N$ of 1100 °C and an even sharper peak centered at 36° for a $T_N$ of 1350 °C. A low temperature nitridation at 850 °C should yield an atomically smoother surface with less protrusions, compared to higher $T_N$ values such as 1100 °C and 1350 °C. This smooth nitridated layer may support the 2D growth of hBN films, since adatom mobility increases for smoother nitriated layer [14].

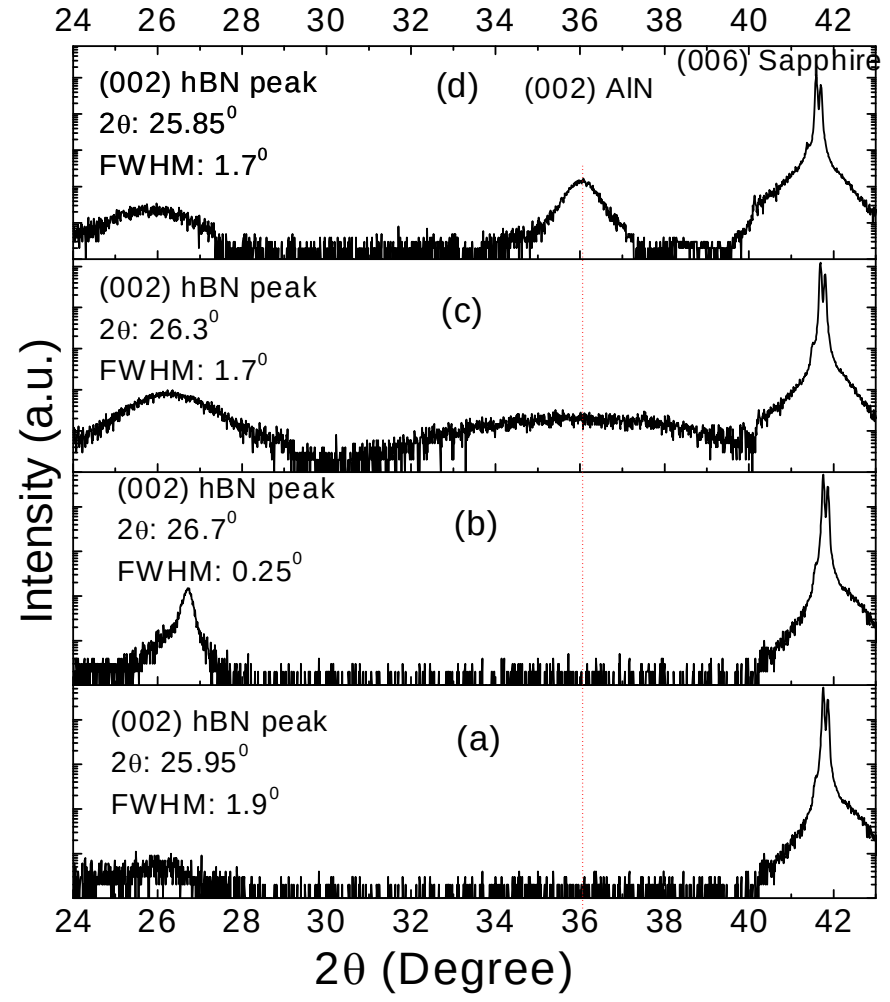

**Figure 1** XRD patterns of samples grown at 1350 °C. (a) without any nitridation step, and with different nitridation temperatures of (b) 850 °C, (c) 1100 °C, and (d) 1350 °C.

Highly crystalline AlN (180 nm thick) grown on (111) Si was also used for hBN growth for this work and the growth recipe used is the same as the one used for hBN growth on nitridated sapphire substrates. No (002) hBN peak was seen in the XRD pattern for this growth (not shown here). Other reported CVD growths of $sp^2$-BN on sapphire incorporated sapphire nitridation at the growth temperature (1200 °C in Ref. [9] and 1500 °C in Ref. [11]) and those films were of similar crystalline quality to the film in this work grown at 1350 °C with sapphire nitridation at 850 °C (indicated by the similar position and FWHM values of the 26.7° peak in the corresponding XRD θ-2θ scans). Epitaxial $sp^2$-BN growth was supported by a thin and amorphous nitridated layer in Ref. [9] and by a thin and strained AlN layer in Ref. [11] (suggested by the (002) AlN peak in the corresponding XRD diffractogram). None of these two works employed different temperatures for nitridation and epitaxial growth, which might have limited their understanding of the suitable nitridated layer quality for the epitaxial growth. This work shows that a much lower $T_N$ of 850 $^oC$ produces a thin and amorphous nitridated layer, which supports epitaxial hBN growth. Higher $T_N$ leads to higher crystallinity of nitridated layer but poorer crystallinity of subsequent hBN. This could originate from increasing protrusion density on the nitridated layer (i.e., rougher surface) with increasing $T_N$, which promotes 3D growth mode of hBN. It indicates that a thin and amorphous nitridated layer serves as the suitable buffer layer to reduce the lattice mismatch between the hBN film and the substrate and/or to improve surface mobility of adatoms, and a low temperature nitridation step can be better for the formation of such buffer layer.

Another crucial process parameter for hBN epitaxy on sapphire is $T_G$. At the early stage of hBN growth by CVD, Y. Kobayashi predicted a necessary $T_G$ of 1500 $^oC$ for single crystal hBN growth on sapphire from the observation of the change of c-lattice constant with increasing temperature [17]. Three films were grown in this work at different $T_G$ values, each with a sapphire nitridation step at 850 $^oC$ for 10 minutes, in order to investigate the effects of $T_G$. Highest crystalline quality hBN was grown at 1350 $^oC$, as shown in Fig. 2 (b). For a lower $T_G$ of 1200 $^oC$, the corresponding film was found to be tBN with a broad (002) hBN peak at 25.85$^o$ (with a FWHM of 2.2$^o$) in an XRD θ-2θ scan, as shown in Fig. 2 (a). A higher $T_G$ of 1500 $^oC$ also resulted in the growth of tBN with (002) hBN peak at 26.3$^o$ with a FWHM of 1.7$^o$, as shown in Fig. 2 (c). A broad peak is found in Fig. 2 (c) around 36$^o$ which might be originated from the improved crystallinity of the initially amorphous nitridated layer (formed by 850 $^oC$ sapphire nitridation) during the 1500 $^oC$ growth; this higher crystalline quality nitridated layer inhibited the epitaxial hBN growth at 1500 $^oC$. These experiments suggest that 1350 $^oC$ is the optimum $T_G$ for hBN growth on a low temperature, thin, and amorphous nitridated layer. The formation of disordered hBN for a $T_G$ of 1200 $^oC$ can be explained by poor adatom mobility on the nitridated layer, since adatom mobility decreases with decreasing temperature. For the 1500 $^oC$ growth, the nitridated layer (formed at a $T_N$ of 850 $^oC$) becomes partially crystalline and may have a rough surface with protrusions. Hence, 1500 $^oC$ growth might also suffer from poor mobility of adatoms on the nitridated later surface, leading to tBN formation.

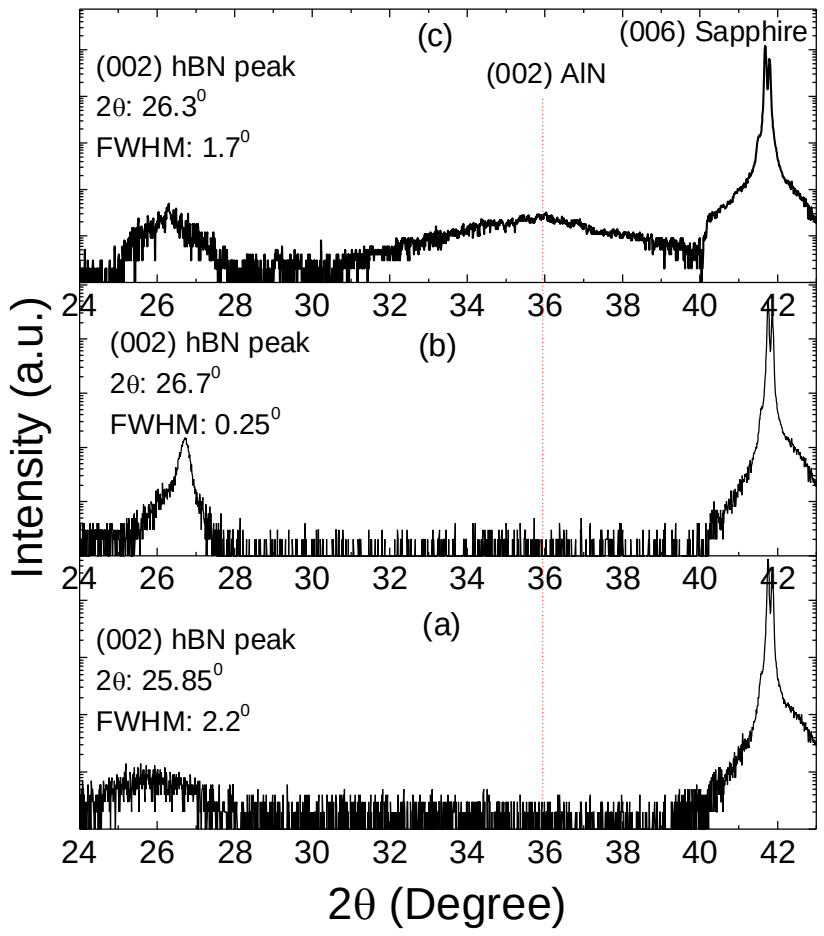

**Figure 2** XRD patterns of samples grown at different growth temperatures of (a) 1200 $^oC$, (b) 1350 $^oC$, and (c) 1500 $^oC$. Sapphire nitridation was done for all three samples at 850 $^oC$.

The phase of the hBN films was characterized by Raman scattering measurements. Figure 3 (a) illustrates Raman spectrum of an hBN film grown on sapphire using the optimum nitridation and growth temperature conditions (i.e., $T_N$ = 850 $^oC$ and $T_G$ = 1350 $^oC$). The hBN film has a thickness of 300 nm. A strong Raman peak is found at 1370.5 cm$^{-1}$ (with a FWHM of 25 cm$^{-1}$), and this Raman active mode can be assigned to the counter phase BN vibrational phase within the BN sheets. This Raman peak position is almost at the same frequency (with broader spectral width) as other reported hBN films [20, 21].

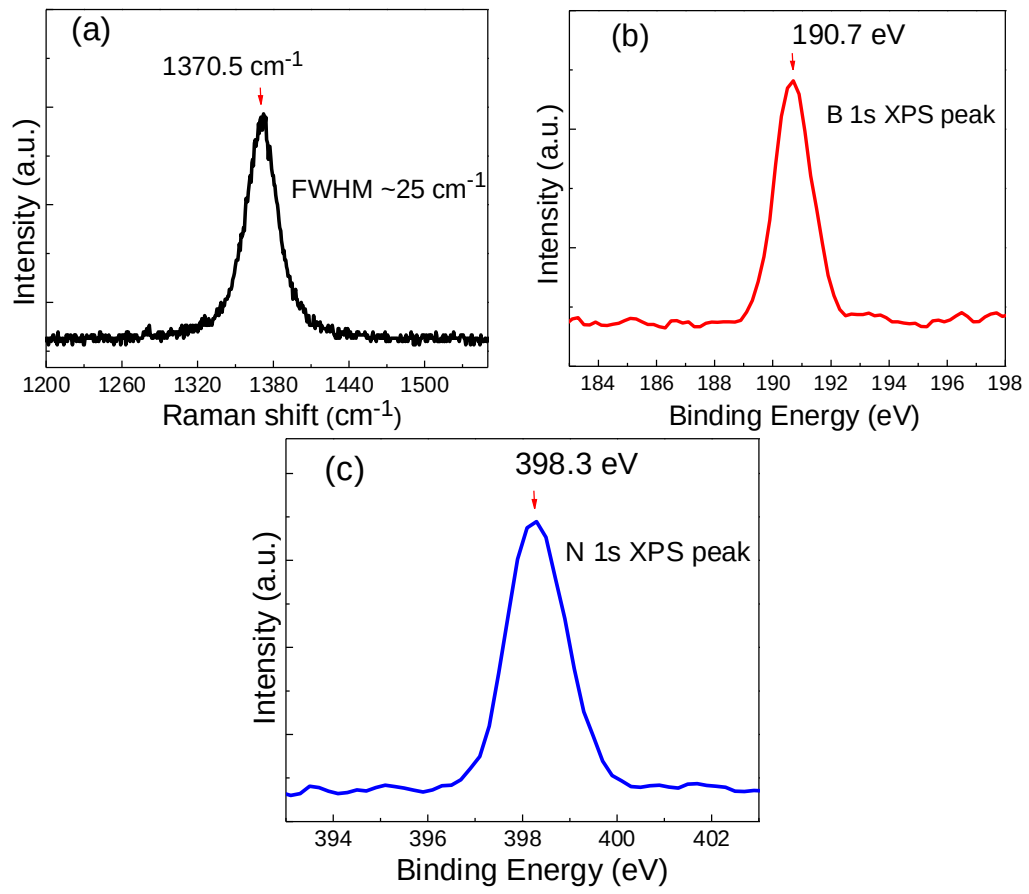

**Figure 3** (a) The Raman spectrum of a 300 nm thick hBN film on sapphire. XPS characterization of the same film: (b) B1s and (c) N1s XPS spectra.

XPS analysis was used as a quantitative spectroscopic technique to determine the elemental composition and analyze chemical bonding states of the hBN films. XPS survey on hBN films demonstrate strong B 1s and N 1s XPS peaks for the B-N bonding state, centered at the binding energy values of 190.7 and 398.3 eV, respectively. These binding energies are similar to the reported literature values [22, 23]. Figure 3 (b) and (c) show these XPS peaks measured for the same 300 nm hBN film. All XPS spectra were shifted with respect to the adventitious C 1s line at 284.8 eV to eliminate the effect of surface charging. Data analysis was done using Multipak software package. Low energy $Ar^+$ ion sputtering was used to remove surface contaminants from the samples. A pass energy of 187.85 eV (with an energy step of 0.5 eV) was employed for initial scans and a pass energy of 23.50 eV (with an energy step of 0.2 eV) was used for high resolution tight scans. Estimated B/N ratio for hBN samples varied from 1.03 to 0.99, and atomic compositions are fairly same across the samples suggesting good sample uniformity.

Optical absorbance and absorption coefficient spectra of the same 300 nm hBN film are shown in Fig. 4 (a). There is a sharp increase in absorbance around photo energy of 5.5 eV (i.e., wavelength of 225 nm), which can be attributed to the near bandgap absorption of the hBN film. Experimental studies [23-26] reported bandgap energies of hBN ranging from 5.6 eV to 6 eV, and both direct and indirect bandgap structure. Recently, Cassabois et al. demonstrated that hBN is an indirect semiconductor of 5.955 eV bandgap from optical spectroscopy measurement [25]. The absorption coefficient of an indirect bandgap semiconductor adheres to the following bandgap equation known as the Tauc's law: $(\alpha E)^{1/2} \propto (E - E_g)$, where α is the absorption coefficient, E is the photon energy, and $E_g$ is the bandgap [23]. Figure 4 (b) shows the Tauc's plot, i.e., $(\alpha E)^{1/2}$ plotted as a function of E for the same hBN film. By extrapolating the linear portion of this curve, an indirect bandgap of 5.47 eV was estimated for the hBN film, which is smaller than indirect bandgap values of hBN reported elsewhere (e.g., 5.955 eV in Ref. [25] and 5.8 eV in Ref. [26]).

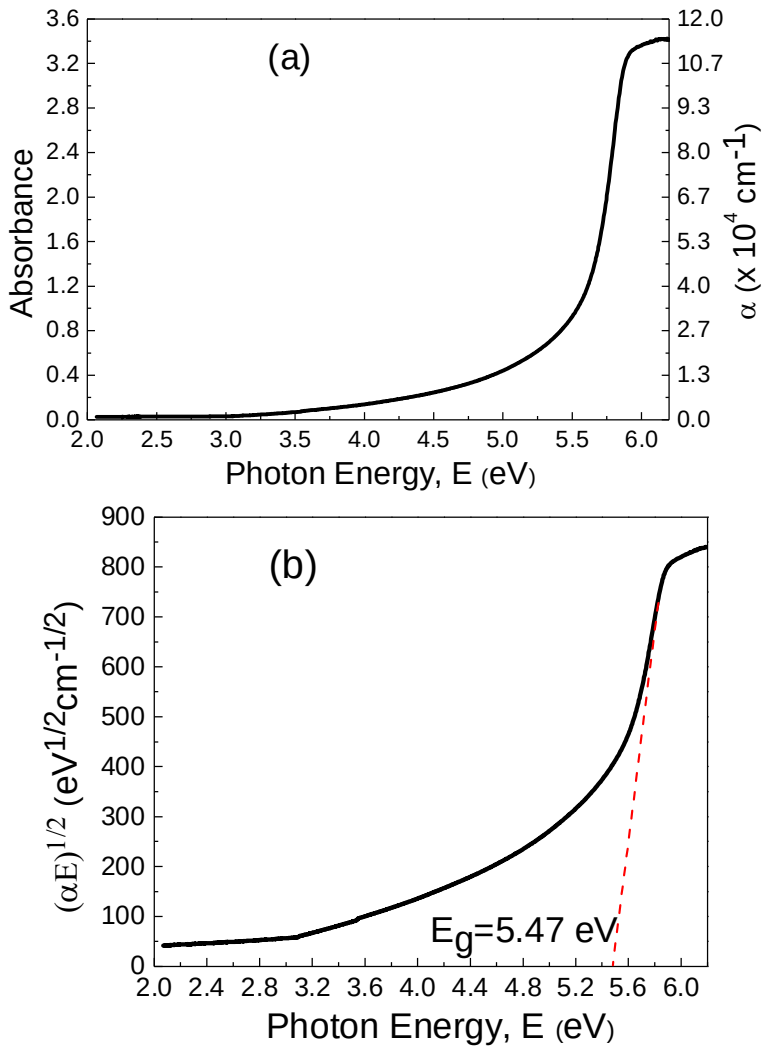

**Figure 4** (a) Absorbance and absorption coefficient (α) spectra of a 300 nm hBN film on sapphire and (b) $(\alpha E)^{1/2}$ versus photon energy for the same hBN film.

**4 Conclusions** Epitaxial growths of hBN films on sapphire substrates were demonstrated at 1350 °C with a low temperature (850 °C) sapphire nitridation step using a cold wall CVD system. Influences of fundamental process parameters such as $T_N$ and $T_G$ were investigated at a reactor pressure of 100 torr and a high V/III ratio of 300. A low temperature (850 °C) sapphire nitridation, known to form a thin amorphous $AlN_xO_{1-x}$ buffer layer, may be suitable for epitaxial growth of hBN at a $T_G$ of 1350 °C. Both higher $T_N$ and higher $T_G$ improved nitridated layer crystallinity and yielded disordered tBN films, while thick high quality AlN was not suitable for even tBN growth. Only a low temperature, thin, and amorphous nitridated layer promoted the nucleation and 2D growth of hBN, which could arise from the nitridated layer being atomically smooth to ensure high adatom mobility and/or the nitridated layer effectively reducing the lattice mismatch between the hBN film and the substrate. hBN films grown at optimum conditions demonstrated excellent properties including the c-lattice constant of 6.66 Å, a characteristic Raman peak at 1370.5 $cm^{-1}$, excellent stoichiometry, and an indirect bandgap of 5.47 eV with high band edge absorption coefficient.

**Acknowledgements** The authors would like to acknowledge the support from the staff of the Analytical Biochemistry Core, Energy Materials and Devices Core Facility, and Nanoscale Characterization Core Facility at Rensselaer Polytechnic Institute. This work was financially supported by the US Department of Homeland Security, Domestic Nuclear Detection Office, under the grants ECCS-1348269 and 2013-DN-077-ER001.

## References

[1] J. Li, R. Dahal, S. Majety, J. Y. Lin, and H. X. Jiang, Nucl. Instrum. Methods Phys. Res. A **654**, 417 (2011).
[2] T.C. Doan, S. Majety, S. Grendadier, J. Li, J.Y. Lin, and H.X. Jiang, Nucl. Instrum. Methods Phys. Res. A **783**, 121 (2015).
[3] O. Tao, C. Yuanping, X. Yuee, Y. Kaike, B. Zhigang, and Z. Jianxin, Nanotechnology **21**, 245701 (2010).
[4] R. Y. Tay, X. Wang, S. H. Tsang, G. C. Loh, R. S. Singh, H. Li, G. Mallick, and E. H. T. Teo, J. Mater. Chem. C **2**, 1650 (2014).
[5] W. Gannett, W. Regan, K. Watanabe, T. Taniguchi, M.F. Crommie, and A. Zettl, App. Phys. Lett. **98**, 242105 (2011).
[6] R. Dahal, K. Ahmed, J. W. Wu, A. Weltz, J.-Q. Lu, Y. Danon, and I. Bhat, Appl. Phys. Express **9**, 065801 (2016).
[7] N. Guo, J. Wei, L. Fan, Y. Jia, D. Liang, H. Zhu, K. Wang, and D. Wu, Nanotechnology **23**, 415605 (2012).
[8] S. Suzuki, R. M. Pallares, C. M. Orofeo, and H. Hibino, J. Vac. Sci. Technol. B **31**, 041804 (2013).
[9] N. Umehara, A. Masuda, T. Shimizu, I. Kuwahara, T. Kouno, H. Kominami, and K. Hara, Jpn. J. Appl. Phys. **55**, 05FD09 (2016).
[10] M. Chubarov, H. Pedersen, H. Högberg, V. Darakchieva, J. Jensen, P. Persson, and A. Henry, Phys. Status Solidi RRL **5**, 397 (2011).
[11] M. Chubarov, H. Pedersen, H. Högberg, J. Jensen, and A. Henry, Cryst. Growth Des. **12**, 3215 (2012).
[12] M. Chubarov, H. Pedersen, H. Högberg, J. Jensen, A. Henry, and Z. Czigany, J. Vac. Sci. Technol. A **33**, 061520 (2015).
[13] D. Skuridina, D.V. Dinh, M. Pristovsek, B. Lacroix, M.-P. Chauvat, P. Ruterana, M. Kneissl, P. Vogt, Appl. Surf. Sci. **307**, 461 (2014).
[14] K. Uchida, A. Watanabe, F. Yano, M. Kouguchi, T. Tanaka, S. Minagawa, J. Appl. Phys. **79**, 3487 (1996).
[15] H. Kawakami, K. Sakurai, K. Tsubouchi, and N. Mikoshiba, Jpn. J. Appl. Phys. **27**, L161 (1988).
[16] J. S. Paek, K. K. Kim, J. M. Lee, D. J. Kim, M. S. Yi, D. Y. Noh, H. G. Kim, and S. J. Park, J. Cryst. Growth **200**, 55 (1999).
[17] Y. Kobayashi and T. Akasaka, J. Cryst. Growth **310**, 5044 (2008).
[18] R. W. Lynch and H. G. J. Drickamer, J. Chem. Phys. **44**, 181 (1966).
[19] A. Yamamoto, M. Tsujino, M. Ohkubo, and A. Hashimoto, J. Cryst. Growth **137**, 415 (1994).
[20] X. K. Cao, B. Clubine, J. H. Edgar, J. Y. Lin, and H. X. Jiang, Appl. Phys. Lett. **103**, 191106 (2013).
[21] Z. Zhao, Z. Yang, Y. Wen, and Y. Wang, J. Am. Ceram. Soc. **94**, 4496 (2011).
[22] L. Wang, B. Wu, L. Jiang, J. Chen, Y. Li, W. Guo, P. Hu, and Y. Liu, Adv. Mater. **27**, 4858 (2015).
[23] R. Y. Tay, S. H. Tsang, M. Loeblein, W. L. Chow, G. C. Loh, J. W. Toh, S. L. Ang, and E. H. T. Teo, Appl. Phys. Lett. **106**, 101901 (2015).
[24] K. Watanabe, T. Taniguchi and H. Kanda, Nat. Mater. **3**, 404 (2004).
[25] G. Cassabois, P. Valvin, B. Gil, Nature Photon. **10**, 262 (2016).
[26] J. H. Edgar, T. B. Hoffman, B. Clubine, M. Currie, X. Z. Du, J. Y. Lin, and H. X. Jiang, J. Cryst. Growth **403**, 110 (2014).